# The Effect of Protein Length on the Ploidy Level and Environmental Tolerance of Organisms

Brian R. Ginn[12]

**Abstract**

This paper summarizes previous work linking protein aggregation to the heterozygosity of organisms. It also cites the literature showing a correlation between species' morphological complexity and the lengths of their proteins. These two findings are combined to form a theory that may potentially explain the ploidy levels of organisms. Organisms can employ heterozygosity to inhibit protein aggregation. Hence, complex organisms tend to be diploid because they tend to synthesize long, aggregation-prone proteins. On the other hand, simple organisms tend to be haploid because they synthesize short proteins that are less prone to aggregation. The theory may also explain ecological trends associated with organisms of different ploidy level. Two mathematical models are also developed that may explain: 1) how protein aggregation results in truncation selection that maintains numerous polymorphisms in natural populations, and 2) the relationship between protein turnover, metabolic efficiency, and heterosis.

1 Department of Crop and Soil Sciences, 3111 Miller Plant Sciences Building, University of Georgia, Athens, GA 30602
2 Email: bginn3@gmail.com

**1.0 Introduction**

The ploidy level of an organism refers to the number of chromosome sets it has. While some species have only one set of chromosomes, others may have two or more sets. Many have argued that higher ploidy levels benefit organisms by masking deleterious mutations (Crow and Kimura, 1965; Charlesworth, 1991; Otto and Goldstein, 1992; Otto and Whitton, 2000). Having two sets of chromosomes, for instance, would mask deleterious recessive alleles and increase the fitness of a diploid organism. This explanation is compatible with the generally accepted partial dominance theory of inbreeding depression. However, the matter is still up for debate (see Gerstein and Otto, 2009 for review). One problem with this perspective is that the "masking effect" that benefits organisms with higher ploidy levels is also what allows the deleterious recessive alleles to persist in the population. The benefits of increased ploidy levels should, therefore, be temporary (see Otto and Whitton (2000) and Otto (2007) for review).

The subsequent sections of this paper will summarize the findings of others, notably Hawkins *et al.* (1986) and Kristensen *et al.* (2002), that have found correlations between heterozygosity and protein metabolism, suggesting a link between heterozygosity and protein aggregation. This paper will also summarize Ginn (2010), which attempted to provide a physical mechanism that could explain these findings. The benefit of the proposed mechanism was that it could explain the geographical distribution of polyploid organisms as well as the correlation between heterozygosity and protein aggregation. This paper will then expand on the theory of Ginn (2010) to provide a potential explanation for the occurrence of haploid and diploid organisms, which will require an overview of the work of previous authors that found morphologically complex organisms tend to synthesize longer proteins than morphologically simple organisms (Patthy, 2003; Tordai et al., 2005; Ekman *et al.*, 2007).

The theory developed in this paper will argue that complex organisms tend to be diploid because they synthesize longer, more aggregation-prone proteins than simple organisms. The theory may help explain the preponderance of simple haploid organisms (such as bryophytes, lichens, and prokaryotes) in extreme environments that promote protein aggregation. Complex organisms suffer from high levels of protein aggregation in these environments, which hinders their ability to migrate into and adapt to these environments unless they are polyploids. In contrast, simple organisms are already pre-adapted to these environments because they experience much lower levels of protein aggregation.

**2.0 Previous Theory**

*2.1 Inbreeding and metabolic efficiency*

Several studies have attempted to correlate the growth rate of organisms to their heterozygosity as measured by allozyme or microsatellite markers (Mitton and Koehn, 1985; Danzmann *et al.*, 1987; Mitton, 1993; Hedgecock *et al.*, 1996; Pogson and Fevolden, 1998; Bayne *et al.*, 1999; Hawkins and Day, 1999; Bayne, 2004; Borrell *et al.*, 2004; Pujolar *et al.*, 2005; Liu *et al.*, 2006; Ketola and Kotiaho, 2009). Some of these studies' authors have argued that their correlations indicate that inbred organisms are less metabolically efficient than outbred organisms (Mitton, 1993; Mitton, 1997; Pogsen and Fevolden, 1998; Borrell *et al.*, 2004). While most studies have focused on the size and mass of the organism, metabolic efficiency can affect other fitness traits. For instance, Gajardo and Beardmore (1989) and Gajardo *et al.* (2001) have shown a positive correlation between heterozygosity and the percentage of female *Artemia* that produce energetically expensive encysted offspring rather than energetically cheaper nauplii. While many studies have concluded that heterozygosity is correlated with the metabolic efficiency of organisms, there is still no consensus view on the underlying mechanism behind this correlation.

The correlation may be explained by protein turnover (Hawkins *et al.*,1986; Hedgecock *et al.*, 1996; Bayne, 2004). Hawkins *et al.* (1986) showed that inbreeding results in higher levels of protein turnover in the blue mussel, *Mytilus edulis*, using $^{15}$N labeled food and allozyme markers. Protein turnover refers to an organism's daily degradation and synthesis of proteins, both of which are energy consuming processes. Therefore, these papers argued, an inbred organism's biomass may be more energetically expensive to sustain than an outbred organism's biomass due to higher levels of protein turnover.

Furthermore, Kristensen *et al.* (2002) and Pedersen *et al.* (2005), using an enzyme-linked immunosorbant assay, found that inbred fruit flies (*Drosophila melangaster* and *Drosophila buzzati*) synthesized more heat shock proteins (*Hsps*) than outbred fruit flies at benign and elevated temperatures. Since *Hsps* are a type of molecular chaperone, proteins that bind to unfolded polypeptide chains and prevent their aggregation, the authors of these papers concluded that inbred fruit flies contain a higher number of unfolded or misfolded polypeptide chains than outbred fruit flies, even at benign temperatures. Kristensen *et al.* (2002) and Kristensen *et al.* (2009) used partial dominance theory to explain their findings. They argued that proteins encoded by deleterious recessive alleles may be less stable, and more prone to aggregation, than the proteins encoded by normal alleles. Consequently, the increased expression of deleterious recessive alleles by inbred organisms may increase their demand for molecular chaperones.

Protein aggregation affects an organism's protein turnover in several ways. Molecular chaperones can bind to unfolded polypeptide chains, preventing their aggregation, or they can tag the polypeptide chains with ubiquitin, thereby marking the polypeptide chains for destruction by the proteasome (Hayes and Dice, 1996; Maurizi, 2002; McClellen *et al.*, 2005). The proteasomal system also degrades proteins after they have aggregated (Dougan *et al.*, 2006; Rubinsztein, 2006; Liberek *et al.*, 2008; Tetzlaff *et al.*, 2008). Finally, protein aggregates may also be degraded via autophagy, whereby aggregated proteins are transported to lysosomes and digested (Kopito, 2000; Garcia-Mata *et al.*, 2002; Kruse *et al.*, 2006; Yorimitsu and Klionsky, 2007). Choe and Strange (2008) observed that half of the genes up-regulated when the nematode *Caenorhabditis elegans* is exposed to aggregate promoting environmental stresses are associated with protein degradation. Especially up-regulated were genes associated with proteasomal and lysosomal degradation. New polypeptide chains will have to be synthesized to take the place of degraded chains, so the correlation between heterozygosity and the expression of *Hsps* observed in Kristensen *et al.* (2002) and Pedersen *et al.* (2005) is directly linked to the correlation between heterozygosity and protein turnover observed in Hawkins *et al.* (1986).

Ginn (2010) provided an explanation for the correlation between heterozygosity and protein turnover/aggregation that differed from the explanation offered by Kristensen *et al.* (2002) and Kristensen *et al.* (2009). According to this alternative theory, heterozygosity dilutes the concentration of unfolded polypeptide chains, thereby favoring protein folding reactions over self-binding reactions. This would mean that inbreeding's effect on protein aggregation is due to the statistical mechanics of protein folding rather than genetics. The theory is summarized below.

*2.2 Model*

All polypeptide chains must fold into their correct conformation in order to be functional. The folding takes time and may be delayed if the folding chain becomes trapped in a metastable intermediate state (Onuchi *et al.*, 1995; Levy *et al.*, 2005; Nevo *et al.*, 2005). Nevertheless, folding proceeds according to a first-order rate law (Keifhaber *et al.*, 1991):

$$\frac{d[N]}{dt} = k_f[U] \qquad (1)$$

where *[N]* is the concentration of native protein, *t* is time, $k_f$ is the rate constant for folding, and *[U]* is the concentration of unfolded polypeptide chains. There is also the potential for an unfolded polypeptide chain to bind with another unfolded chain and form a soluble oligomer (Silow and Oliveberg, 1997; Bitan *et al.*, 2001; Kayed *et al.*, 2003; Kayed *et al.*, 2004; Cleary *et al.*, 2005; Haass and Selkoe, 2007; Viera *et al.*, 2007; Wei *et al.*, 2007). Soluble oligomers can then bind with additional unfolded chains and eventually become a solid protein aggregate. The process of protein aggregation is highly specific in that protein aggregates are highly enriched with a single protein species, even when two or more polypeptides are aggregating simultaneously (London *et al.*, 1974; Speed *et al.*, 1996; Kopito, 2000; Rajan *et al.*, 2001; Morell *et al.*, 2008). The process may be so specific that small differences in amino acid sequence can inhibit co-aggregation of different polypeptide chains. For example, O'Nuallain *et al.* (2004) found that a single point mutation can prevent amyloid fibrils from co-aggregating. The specificity of protein aggregation implies that the formation of soluble oligomers proceeds as a second order reaction (Keifhaber *et al.*, 1991; Bitan *et al.*, 2001; Zhdanov and Kasemo, 2004; Zhu *et al.*, 2010):

$$\frac{d[O]}{dt} = k_b[U]^2 \qquad (2)$$

where *[O]* is the concentration of the soluble oligomer and $k_b$ is the rate constant for self-binding.

The formation of soluble oligomers and solid protein aggregates is detrimental for two reasons. First, as shown in Figure 1, the formation of soluble oligomers and protein aggregates competes with the proper folding of a protein (Keifhaber *et al.*, 1991). A protein's folding efficiency decreases when more unfolded polypeptide chains bind to each other. Second, soluble oligomers and solid protein aggregates are cytotoxic species that have been associated with several disorders (Haas and Selkoe, 2007; Viera *et al.*, 2007).

A comparison of Equations 1 and 2 reveals that the rate of protein self-binding is more dependent upon the concentration of unfolded polypeptide chains than is the rate of protein folding. For example, an organism that is homozygous at a gene locus will synthesize only one type of polypeptide chain, called *A*, which will fold at the rate:

$$\frac{d[N_A]}{dt} = k_{fA}[U_A] \tag{3}$$

where $[N_A]$ is the concentration of native protein $A$, $k_{fA}$ is the rate constant for the folding of $A$, and $[U_A]$ is the concentration of unfolded polypeptide chain $A$. The rate of $A$'s self-binding to form a soluble oligomer will be:

$$\frac{d[O_A]}{dt} = k_{bA}[U_A]^2 \tag{4}$$

On the other hand, an organism that is heterozygous at the same gene locus will synthesize two different types of polypeptide chains, $A$ and $A^\ddagger$. The $A^\ddagger$ polypeptide chains will be synthesized in place of the $A$ polypeptide chains, so at any given time a heterozygous organism will have half the concentration of $A$ as a homozygous organism. Likewise, a heterozygous organism will have half the concentration of $A^\ddagger$ as a homozygous organism. The overall rate of folding for the heterozygous organism will be:

$$\frac{d[N]}{dt} = 0.5k_{fA}[U_A] + 0.5k_{fA^\ddagger}[U_{A^\ddagger}] \tag{5}$$

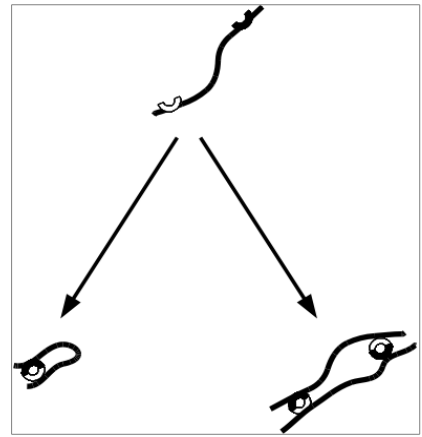

Figure 1. An unfolded polypeptide chain may either fold into its native conformation or bind to another unfolded polypeptide chain. The two reactions compete against each other, and the relative rates of each reaction will determine the folding efficiency of the polypeptide chain. The rings on the polypeptide chain represent binding sites that stabilize the folded protein or soluble oligomer.

where $[U_A]$ and $[U_A^\ddagger]$ are the concentrations of polypeptide chains $A$ and $A^\ddagger$ if the organism were homozygous for either polypeptide chain. If the rate constants of $A$ and $A^\ddagger$ folding are comparable, then the overall rate of polypeptide chain folding in a heterozygous organism will be approximately the same as in a homozygous organism. Equation 5 can be expressed more generally as:

$$\frac{d[N]}{dt} = \sum r_i k_{fi}[U_i] \tag{6}$$

where $r_i$ is the concentration of polypeptide chain $i$ in a heterozygous organism divided by the concentration of $i$ in a homozygous organism, $k_{fi}$ is the folding rate constant for $i$, and $[U_i]$ is the concentration of unfolded chain $i$ in a homozygous organism.

The kinetics of protein self-binding are going to be considerably different. In the case of a heterozygous organism, the overall rate of protein self binding will be:

$$\frac{d[O]}{dt} = 0.25k_{bA}[U_A]^2 + 0.25k_{bA^\ddagger}[U_{A^\ddagger}]^2 \tag{7}$$

If the self-binding rate constants of $A$ and $A^{\ddagger}$ are comparable, then the rate of protein self-binding in a heterozygous organism will be approximately half that of a homozygous organism. Equation 7 can be expressed more generally as:

$$\frac{d[O]}{dt} = \sum r_i^2 k_{bi} [U_i]^2 \tag{8}$$

Equations 3-8 show that diluting the concentration of an unfolded polypeptide chain shifts the competition between protein folding and self-binding in favor of folding. Thus, heterozygosity increases the folding efficiency of unfolded polypeptide chains simply by diluting their concentrations.

Another way to think of the influence that heterozygosity has on protein self-binding is to consider the number of collisions that will occur in a given time period. In a homozygous organism, all of the collisions will be $A$-$A$ collisions because $A$ is the only polypeptide chain that is present. In contrast, both $A$ and $A^{\ddagger}$ are present in a heterozygous organism, so 25% of the collisions in the heterozygous organism will be $A$-$A$ collisions, 25% of the collisions will be $A^{\ddagger}$-$A^{\ddagger}$ collisions, and 50% of the collisions will be $A$-$A^{\ddagger}$ collisions. Since only $A$-$A$ and $A^{\ddagger}$-$A^{\ddagger}$ collisions will result in self-binding reactions, the heterozygous organism will have half as many self-binding reactions as the homozygous organism within the given time period. Each variety of polypeptide chain buffers the self-binding reaction of the other variety. The critical assumption is that protein self-binding is highly specific, which is corroborated by the research papers cited above.

The theory presented in Ginn (2010) provided a statistical mechanical explanation for the correlation between heterozygosity and molecular chaperone expression. Organisms may increase their expression of molecular chaperones as a response to the buildup of soluble oligomers within their cytosol. Consequently, inbred organisms will synthesize more molecular chaperones than outbred organisms because their proteins have lower folding efficiencies. This in turn will cause inbred organisms to have higher rates of protein turnover and lower metabolic efficiencies than outbred organisms. Thus, heterozygosity benefits organisms even in the absence of deleterious recessive alleles. The advantage of this explanation over the genetic explanation will hopefully be demonstrated in the following sections of this paper. More details of the theory are provided in Appendices 1 and 2. Appendix 1 covers truncation selection for heterozygosity and describes in more detail how heterozygosity could lead to greater metabolic efficiency. Appendix 2 covers the relationship between heterozygosity and chemical potential.

*2.3 Polyploidy*

If synthesizing two varieties of protein can inhibit protein self-binding, then synthesizing three or more varieties should inhibit protein self-binding even more. Equation 8 predicts that polyploid organisms should have lower rates of protein aggregation than diploid organisms. Furthermore, most polyploids are hybrid species, and are heterozygous at more gene loci than diploids (Whitton and Otto, 2000). Unfortunately, there is no experimental data that compares protein aggregation rates in polyploids and diploids, but experiments have shown that triploid mollusks have greater metabolic efficiencies than their diploid relatives (Hedgecock *et al.*, 1996; Hawkins *et al.*, 2000). No attempt has been made to compare the protein turnover or *Hsp* expression levels of polyploid and diploid organisms. However, the environments in which polyploid organisms grow provide evidence that polyploid organisms have reduced rates of protein aggregation compared to diploids.

Several environmental variables promote protein aggregation. Among them are freezing, desiccation, and salinity (Goyal *et al.*, 2005; Choe and Stange, 2008). All of these conditions limit the amount of liquid water within the cell and promote molecular crowding. Furthermore, organisms respond to freezing, desiccation, and high salinity by producing compatible solutes, such as polyols, which promote further crowding and protein aggregation (Ellis, 2001; Smallwood and Bowles, 2002; Chebotareva *et al.*, 2003; Ellis and Minton, 2006). Ellis and Minton (2006) proposed that the enhanced protein aggregation rates caused by molecular crowding can be appropriately modeled simply by multiplying the self-binding constant by a crowding coefficient:

$$k_b = k_b^\circ \Gamma \qquad (9)$$

where $k_b$ is the observed self-binding constant, $k_b^\circ$ is the self-binding constant in the absence of molecular crowding, and $\Gamma$ is the crowding coefficient.

Organisms respond to freezing, desiccation, and high salinity by activating defense mechanisms that inhibit protein aggregation, such as *Hsps* and late embryonic associated (*LEA*) proteins (Goyal *et al.*, 2005; Chakrabortee *et al.*, 2006; Rinehart *et al.*, 2006; Hundertmark and Hincha, 2008; Wang *et al.*, 2009; Boucher *et al.*, 2010). Thus, heterozygosity would be most advantageous in frozen, arid, and hypersaline environments, and polyploid organisms should be positively associated with these environments.

The occurrence of polyploid plants at high latitudes and altitudes was first observed in the 1940's (Stebbins, 1950; Stebbins, 1984), and recent research has confirmed that polyploid plants and animals frequently occur in frozen environments (Adamowicz *et al.*, 2002; Brockmann *et al.*, 2004; Lundmarke and Saura, 2006; Aguilera *et al.*, 2007; Otto *et al.*, 2007). Brockmann *et al.* (2004) analyzed compiled data from the *Pan-Arctic Flora (PAF) Checklist* (Elven *et al.*, 2003) and found that 73.7% of arctic plants are polyploid. Plants in the most northerly arctic zone were hexaploid on average. In addition, 39.2% of species within this zone were 7-ploid or higher, and 17.8% of the species were 9-ploid or higher. However, since these plants reproduce primarily through self-fertilization, the heterozygosity of these plants is about half what their ploidy level would indicate. For example, a tetraploid selfing plant has the same heterozygosity as a non-selfing diploid plant and a hexaploid selfing plant has the same heterozygosity as a non-selfing triploid plant.

Polyploid plants are also positively associated with arid zones and deserts (Spellenberg, 1981; Rossi *et al.*, 1999; Hunter *et al.*, 2001; Pannell *et al.*, 2004; Joly *et al.*, 2006; Schuettpelz *et al.*, 2008). Most resurrection plants, which grow in deserts and are capable of withstanding very high levels of desiccation, are polyploid (Bartels and Salamini, 2001; Rodriguez *et al.*, 2010). Several studies have shown that a plant's drought tolerance increases with its ploidy level (Al-Hakimi *et al.*, 1998; Xiong *et al.*, 2006). For example, Ramsey (2011) compared the drought tolerance of hexaploid and tetraploid individuals belonging to *Achillea borealis* and found the hexaploids were more tolerant. Furthermore, Ramsey's analysis of neo-hexaploid *A. borealis* individuals showed that a third of the the drought tolerance was achieved via genome duplication. Polyploid animals are also positively associated with arid zones. For example, polyploid lizards are found in deserts (Tocidlowski *et al.*, 2001; Kearney, 2003) and so is the only known polyploid mammal (Svartman *et al.*, 2005; Gallardo *et al.*, 2006).

Evidence for a link between polyploidy and salinity is unfortunately tenuous. However, many species change their expression of *Hsps* in both hypersaline and hyposaline environments (Chang, 2005; Downs *et al.*, 2009; Tine *et al.*, 2010; Monari *et al.*, 2011), so a connection between salinity and polyploidy is to be expected. Polyploid plants appear to have greater tolerance to salt stress than their diploid relatives (Tal and Gardi, 1976; Shannon and Greive, 1999; Ashraf *et al.*, 2001; Kumar *et al.*, 2009). Also, several species of polyploid brine shrimp, *Artemia*, have been identified (Browne and Bowen, 1991; Amat *et al.*, 2007), though their distribution is controlled primarily by latitude (Zhang and Lefcort, 1991). Polyploid *Artemia* actually tolerate lower salinities than diploids, which is in accord with hyposalinity inducing expression of *Hsps* (Zhang and King, 1993; Agh et al., 2007; Chang, 2005; Monari *et al.*, 2011).

The geographical distribution of polyploid organisms suggests that they have an advantage in environments that promote protein aggregation. The statistical mechanical theory of heterosis provides a simple explanation for the benefit of heterozygosity in these environments. The partial dominance theory of heterosis struggles to explain these geographical trends because polyploidy cannot provide a long-term advantage by masking deleterious mutations (Whitton and Otto, 2000). Polyploidy can temporarily mask the genetic load within a population, but the masking effect allows deleterious mutations to accumulate within the population over time. If polyploid individuals have a higher mutation rate than diploid individuals (because polyploids have more gene loci where a mutation can occur), then the genetic load of polyploid populations will eventually be greater than the genetic load of otherwise equivalent diploid populations.

**3.0 Protein length and the haploid/diploid transition**

Ginn (2010) attempted to relate the heterozygosity of organisms to their levels of protein aggregation and protein turnover. The resulting theory was then used to explain the geographical distributions of polyploid organisms. However, it never addressed why some organisms are haploid whereas others are diploid. In this paper, a new theory will be developed that might explain the occurrence of haploid and diploid organisms. The new theory may also explain why some organisms, such as bacteria and archaea, do not take advantage of heterozygosity, even in extreme environments that promote protein aggregation. According to the new theory, the lengths of proteins synthesized by individuals will determine the extent that heterozygosity will improve their fitness. Therefore, organisms that synthesize smaller (less aggregation-prone) proteins tend to be haploid whereas organisms that synthesize longer (more aggregation prone) proteins tend to be diploid. Furthermore, organisms that synthesize shorter proteins can tolerate harsh environments better than organisms that synthesize longer proteins. The new theory contrasts with Ginn (2010) because it focuses on the impact protein length has on ploidy level whereas Ginn (2010) focused on the impact environmental stresses had on ploidy level.

*3.1 Theory*

Many proteins can be divided into parts called domains. A domain is a sequence of amino acids that, if separated from the rest of the polypeptide chain, would still fold into its proper conformation and function as a normal protein. A multi-domain protein can be considered a string of proteins that are joined together. In fact, proteins that exist separately in some species may be found as parts of multi-domain proteins in other species, a phenomena called "domain accretion" (Koonin *et al.*, 2002; Basu *et al.*, 2009).

The length of a polypeptide chain should affect the competition between its folding and self-binding reactions. Long polypeptide chains take longer to fold into their native conformations than short chains because they have to form more chemical bonds while folding, and because they can assume more possible conformations before finding their native one (e.g. Ivanokov *et al.*, 2003). As a consequence, long polypeptide chains should have lower folding rate constants, $k_f$ (Equation 1), than short chains. On the other hand, long polypeptide chains have a great potential for self-binding when they approach each other because self-binding can occur at many sites along the unfolded polypeptide chain (e.g. Andersen *et al.*, 2010). Therefore, long polypeptide chains should have higher self-binding constants, $k_b$ (Equation 2), than short chains. As a result of these dynamics, long polypeptide chains should have higher self-binding rates and lower folding efficiencies than short polypeptide chains. Hence, organisms that synthesize long, multi-domain proteins benefit from heterozygosity and should tend to be diploid. Conversely, organisms that synthesize short, less aggregation-prone proteins benefit less from heterozygosity and should tend to be haploid.

The theory developed in this section predicts that diploid organisms synthesize longer proteins, on average, than haploid organisms. The following sections will cite data that support this prediction. They will also cite studies that report the relative stress tolerances and geographical distributions of haploid and diploid organisms because the studies provide circumstantial evidence for the prediction. Haploid organisms should be able to tolerate stresses that promote protein aggregation if they synthesize short proteins, so they should be pre-adapted to stressful environments, which would enable them to migrate into and adapt to such environments. In contrast, diploid organisms should be less tolerant of these kinds of stresses if they synthesize long proteins, so they should struggle to survive in harsh environments. This is true because heterozygosity can only lower, not completely inhibit, the aggregation of long proteins. Therefore, harsh environments should be dominated by haploid organisms, and diploid organisms should be restricted to benign environments.

*3.2 Protein length and prokaryotes*

Prokaryotes, which are overwhelmingly haploid, produce shorter proteins than eukaryotes. About 65% of prokaryote proteins are multi-domain whereas 80% of eukaryote proteins are multi-domain (Apic *et al.*, 2001). Furthermore, the median length of eukaryote proteins is 50% longer than the median length of prokaryote proteins (Brocchieri and Karlin, 2005). These differences have caused prokaryotes and eukaryotes to process their proteins differently. For example, most protein folding is post-translational in prokaryotes but co-translational in eukaryotes (Netzer and Hartl, 1997). Eukaryotes also have more complex chaperone systems that assist with the folding of nascent proteins (Albanèse *et al.*, 2006).

Siller *et al.* (2010) used genetically modified *Escherichia coli* to demonstrate that the proteins of eukaryotes are more aggregation prone than the proteins of prokaryotes. *E. coli* typically translate proteins at a rate of 10-20 amino acids per second whereas eukaryotes typically translate proteins at a rate of 3-8 amino acids per second. Siller *et al.* (2010) used a mutant strain of *E. coli* that translated proteins at rates typical of eukaryotes to examine the extent of aggregation of recombinant proteins taken from *Saccharomyces cerevisae*. Extensive protein aggregation occurred in *E. coli* strains that synthesized the recombinant proteins at wild-type rates, but much less protein aggregation occurred in the *E. coli* strains that synthesized the recombinant proteins at rates typical of eukaryotes. They concluded from their data that *S. cerevisae* proteins fold more slowly than *E. coli* proteins and are more prone to aggregation. Since proteins begin to fold before they are completely synthesized, a slower rate of translation will allow more folding to occur before the nascent polypeptide chain is released from its ribosome. Thus, they argued, eukaryotes need to translate proteins at slower rates than prokaryotes because their proteins are longer, more susceptible to aggregation, and need more time for co-translational folding.

The relatively small size of prokaryote proteins can explain why they are haploid. The extent of protein self-binding is too small for diploid bacteria to have a significant heterozygous advantage over haploid bacteria. The small size of prokaryote proteins can also explain why they can survive in extreme environments without being heterozygous. Koonin *et al.* (2002) and Brocchieri and Karlin (2005) both proposed that the high frequency of thermophiles within the Domain Archaea can be explained by the relatively small size of their proteins. However, the idea can be extended to all prokaryotes that live in extreme environments. Thermophiles can be found among both the Bacteria and the Archaea. Additionally, both domains of life contain species that are able to grow at extremely high salinities (Kunte *et al.*, 2002), withstand desiccation (Potts, 1994; Alpert, 2006), or grow in freezing conditions (Russell, 1998). Not all prokaryotes can grow in extreme environments because specific adaptations, such as those promoting membrane integrity and DNA stability, are required. However, prokaryotes may be pre-adapted to these environments because their small proteins are less aggregation prone and, therefore, more resilient to environmental stresses.

*3.3 Protein length and organism complexity*

Several studies have shown a correlation between organism complexity and protein complexity (Patthy, 2003; Tordai *et al.*, 2005; Ekman *et al.*, 2007). In general, organisms with greater morphological complexity have a higher proportion of multi-domain proteins than simple organisms. Tordai *et al.* (2005) used data from the UniProt Knowledgebase to create databases of proteins synthesized by 2 bacteria, 3 archaea, 1 protist, 1 plant, 2 fungi, and 3 animals. They estimated the relative number of multi-domain proteins in each of these groups by determining the percentage of proteins that had two or more domains annotated in the Pfam-A database. The results, shown in Table 1, show that the proportion of multi-domain proteins decrease in order of animals > plants > fungi ~ protist > bacteria > archaea. 39% of animal proteins contain two or more Pfam-A domains while only 23% of archaea proteins have two or more Pfam-A domains.

Table 1

| Tordai et al., 2005 | | Ekman et al., 2007 | | |
|---|---|---|---|---|
| Category | Multidomain proteins (% of proteins) | Category | Multidomain architectures (% of domain architecture) | Species |
| Bacteria | 27% | Animals | 49% | *Homo sapiens* |
| Archaea | 23% | | 49% | *Mus musculus* |
| Protozoa | 32% | | 49% | *Rattus norvegicus* |
| Plants | 35% | | 49% | *Gallus gallus* |
| Fungi | 32% | | 49% | *Danio rerio* |
| Metazoa | 39% | | 44% | *Drosophila melanogaster* |
| | | | 44% | *Caenorhabditis elegans* |
| | | Plants | 37% | *Arabidopsis thaliana* |
| | | | 35% | *Oryza sativa* |
| | | Fungi | 33% | *Saccharomyces cerevisiae* |
| | | | 33% | *Schizosaccharomyces pombe* |
| | | Eukarya | 53% | |
| | | Bacteria | 29% | |
| | | Archaea | 29% | |

Ekman *et al.* (2007) showed a similar link between morphological complexity and protein complexity. They defined the domain architecture (DA) of a protein as its sequence of domains. If the sequence contained only one domain, then the protein had a single domain architecture (SDA). If the sequence contained multiple domains, then the protein had a multi-domain architecture (MDA). Ekman *et al.* (2007) created databases of proteins from 7 animals, 2 plants, 2 fungi, 7 bacteria, and 7 archaea and determined the percentages of each groups' DAs that were MDAs. The result, shown in Table 1, was that the proportion of MDAs decrease in order of animals > plants > fungi > bacteria ~ archaea. Thus, the results of Ekman *et al.* (2007) are in agreement with Tordai *et al.* (2005).

The trends observed in both Tordai *et al.* (2005) and Ekman *et al.* (2007) correspond to the ploidy levels of the different taxa of organisms. The prokaryotes have the lowest percentage of multi-domain proteins and are haploid. Animals have the highest percentage of multi-domain proteins and are diploid. Plants, fungi, and protists have intermediate percentages of multi-domain proteins and are either haploid or diploid, depending on the species. Wang *et al.* (2005) provided another line of evidence for the relative complexity of animal proteins. They found that proteins shared by *S.*

*cerevisiae*, *D. melangster*, and *Homo sapiens* are similar in length. However, proteins found in *D. melangster* and *H. sapiens,* but not in *S. cerevisiae,* are on average 22% longer than proteins shared by all three species. Therefore, a likely reason why all animal species are (at least) diploid is that animals have the longest proteins of all organisms.

Two additional observations support the theory that diploidy benefits animals by suppressing the aggregation of multi-domain proteins. First, many studies have observed that animals are more sensitive to the stresses of extreme environments than less complex organisms (Alpert, 2006; Kranner *et al.,* 2008). Spider-mites, which possess diploid females and haploid males, provide another interesting example of the theory. Several studies have found that male spider-mites are less tolerant of environmental stresses than females, and that female spider-mites change the sex ratio of their offspring to favor females under stressful conditions (Veerman, 1985; Roy et al., 2003). Again, these findings support the theory because male spider-mites should be less tolerant of stressful environments since they are not heterozygous.

Tan *et al.* (2005) provided a potential explanation for the correlation between organism complexity and protein length. They argued that complex organisms have more complex protein interaction networks than simpler organisms, so their proteins should have more interaction partners, on average. Proteins with many interaction partners should be longer than proteins with few interaction partners because they need many domains to facilitate their various interactions. To test their hypothesis, they analyzed protein sequences from the SWISS-PROT database and found that the lengths of proteins correlated with the number of their interaction partners.

 A simple picture emerges from the studies cited in this section. Multi-domain proteins are required by the protein interaction networks that make organism complexity possible, and complex organisms benefit from heterozygosity because it inhibits the aggregation of multi-domain proteins.

*3.4 Protein length and plants*

Plants typically alternate between a haploid gametophyte generation and a diploid sporophyte generation. However, as shown in Figure 2, the dominant generation varies between species, which can be divided into three categories: bryophytes, ferns, and spermatophytes. Bryophytes (mostly mosses) have a dominant gametophyte generation that reproduce sexually to produce short-lived sporophytes. The sporophytes grow out of their gametophyte parents, and depend on them for sustenance, until they produce haploid spores that germinate to form new gametophyte plants. In contrast, fern gametophytes and sporophytes can exist independently of each other. Fern sporophytes grow out of their gametophyte parents, but they are self-sustaining. In fact, the gametophytes typically die shortly after fertilization. Spermatophytes (seed plants) have a dominant sporophyte generation that encompasses almost the entire life-span of the organism. The gametophyte generation is retained only in the gametes. Thus, plants may sit on the boundary of the haploid-diploid transition, with the haploid-dominant bryophytes on one side of the boundary and the diploid-dominant spermatophytes on the other side of the boundary.

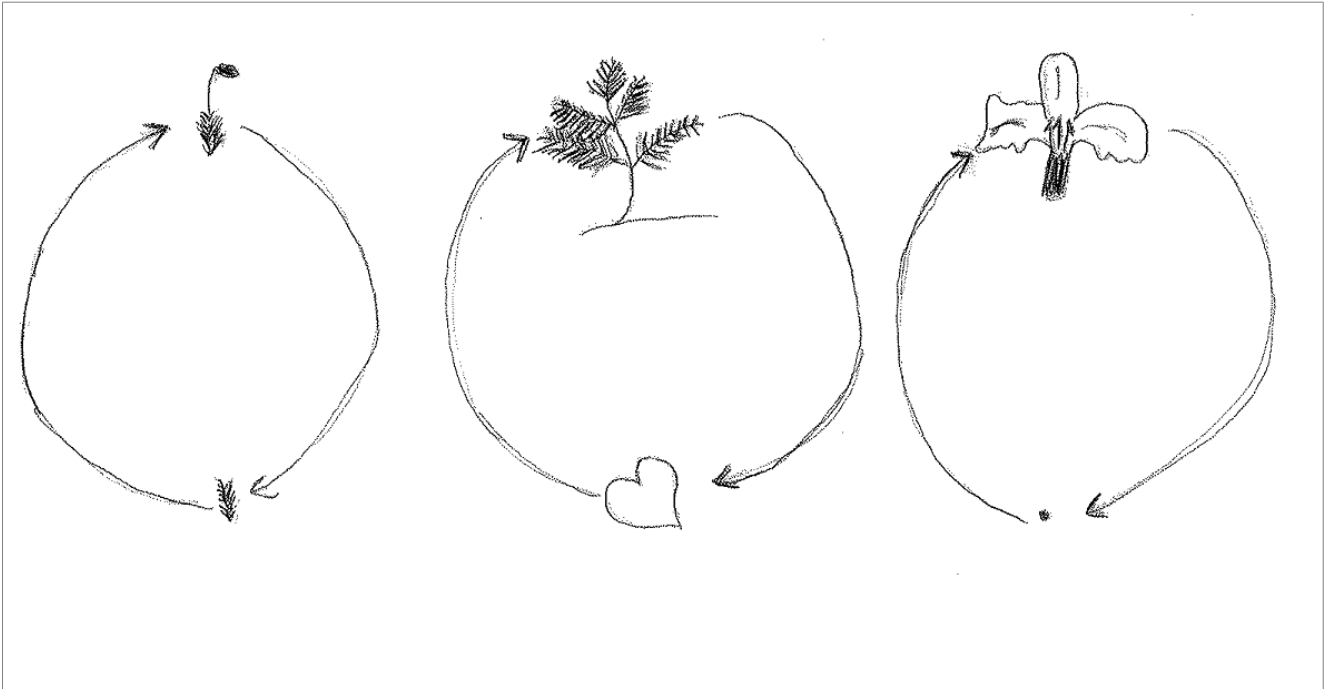

Figure 2. Three different life-cycles found in plants. All of the top stages are diploid and the bottom stages are haploid. The left cycle is typical of bryophytes. A haploid gametophyte plant gives rise to a temporary sporophyte structure. The center cycle is typical of ferns. Simple gametophyte plants alternate with complex sporophyte plants. The right cycle is typical of spermatophytes. A diploid sporophyte plant produces haploid gametes.

Once again, there is a relationship between the complexity of the species and its ploidy level. The haploid-dominant bryophytes are relatively simple plants, typically 2 cm tall and one cell thick. The diploid-dominant spermatophytes are complex and include flowering plants. The ferns alternate between a haploid generation that is simple, resembling bryophytes, and a diploid generation that is significantly larger and more complex. The studies of Tordai *et al.* (2005) and Ekman *et al.* (2007) show that the percentage of multi-domain proteins in plants is the second highest among their categories of life, slightly higher than fungi. However, their studies only included angiosperms and not bryophytes or ferns, which provides an opportunity for a prediction. The theory presented in this paper predicts that bryophytes should have smaller percentages of multi-domain proteins (comparable to the percentages seen in fungi) than spermatophytes.

Support for the prediction comes from the relative stress tolerances of each plant division. Bryophytes are much more tolerant of freezing, desiccation, and salinity stresses than spermatophytes (Alpert, 2000; Oliver *et al.*, 2005; Wang *et al.*, 2009). Their ability to tolerate such stresses is comparable to lichens, and they can be found, along with lichens, in extremely cold and arid environments not inhabited by more complex plants (Longton, 1988; Alpert, 2006; Procter and Tuba, 2002; Kranner *et al.*, 2008). The desiccation tolerance of ferns is more complex. Fern sporophytes are comparable to spermatophytes in their ability to tolerate desiccation, but fern gametophytes are comparable to bryophytes (Watkins *et al.*, 2007; Hietz, 2010). In fact, some tropical fern species have lost the sporophyte stage of their life cycle and now exist as asexually reproducing gametophytes, which allows them to live in colder and drier habitats than their sporophyte-producing relatives (Farrar, 1978; Farrar, 1990). Fern gametophytes and sporophytes probably express different, but overlapping, sets of proteins. The proteins expressed by fern gametophytes probably have similar lengths to the proteins expressed by bryophytes, while the proteins expressed by fern sporophytes probably have similar lengths to the proteins expressed by spermatophytes. This would explain the relative order of plant stress tolerances: bryophyte ~ fern gametophyte > spermatophyte ~ fern sporophyte. The trend corresponds to the relative complexity of the plant divisions and to the relative abundances of multi-domain proteins they are expected to express.

*3.5 Protein length and fungi*

Fungi also exhibit associations between their ploidy levels and environmental tolerances. 98% of all fungi species are either ascomycetes or basidiomycetes (Stajich *et al.*, 2009; HaiYing *et al.*, 2010). Both phyla belong to the Dikarya sub-kingdom and are characterized by a dikaryotic stage in their life-cycle. They germinate from spores as haploid hyphae and mate by exchanging nuclei with nearby hyphae of opposite mating-type. However, the haploid nuclei do not fuse together to form a diploid nucleus. Instead, they remain separate, joined together by a protein clamp connection (Kuess, 2000; Stajich *et al.*, 2009). The dikaryotic state is not truly diploid, but the organism is still heterozygous. Karyogamy (fusion of the nuclei) usually occurs immediately prior to meiosis and the formation of sexual spores. Ascomycetes and basidiomycetes differ substantially in the lengths of the dikaryotic stages of their life-cycles. Ascomycetes primarily grow as haploid hyphae and only produce a temporary dikaryotic fruiting-body, called an "ascocarp." Basidiomyctes, on the other hand, are only temporarily haploid and spend most of their life-cycle as either dikaryotic hyphae or dikaryotic fruiting-bodies, called "basidiocarps." Thus, ascomycetes are haploid-dominant and basidiomycetes are dikaryote-dominant (Küess, 2000; Stajich *et al.*, 2009 ).

Based on the life-cycle differences between the ascomycetes and basidiomycetes, the theory presented in this paper predicts that the basidiomycetes synthesize longer proteins, on average, than the ascomycetes. If this is true, then the theory also predicts that ascomycetes can better tolerate environmental stresses than basidiomycetes. Ascomycetes should be pre-adapted to environments that promote protein aggregation while basidiomycetes should not. In fact, ascomycetes are capable of withstanding many stresses, especially in the lichenized form. Lichens, 98% of which are ascomyctes, are found in many deserts and polar environments and are sometimes the primary producers in such environments (Hawksworth, 1988; Longton, 1988; Kranner *et al.*, 2008). The relative abundances of basidiomycetes and ascomycetes in harsh environments has not been studied extensively, but Schadt et al. (2003) has shown that approximately 90% of fungal species in tundra environments are ascomycetes and approximately 10% are basidiomyctes. This study did not distinguish between lichenized and non-lichenized fungi. However, the relative diversities of ascomycetes and basidiomycetes would suggest that ascomycotes are more capable of surviving in the tough conditions of tundra environments.

*3.6 Conclusion*

This paper has proposed that the ploidy level of organisms can be explained by the lengths of the proteins they produce. Longer proteins that are made of many domains are more prone to self-binding than shorter proteins, so complex organisms, which tend to produce long proteins, benefit more from heterozygosity (and diploidy) than simple organisms, which tend to produce shorter proteins. Additionally, simple organisms should be pre-adapted to harsh environments that promote protein aggregation because they produce short proteins. In contrast, complex organisms often survive in such environments as polyploids, which may have three or more alleles at any given gene locus. Therefore, freezing, arid, and hypersaline environments are expected to be dominated by simple organisms and by polyploid complex organisms. Interestingly, ferns have taken both approaches. Some fern species that have migrated into colder, drier environments have lost the complex sporophyte stage of their life-cycle and live as simple gametophytes (Farrar, 1990). Other fern species have retained their complex sporophyte stage, but are polyploid (Schuettpelz *et al.*, 2008). The power of this paper's theory is that it can explain the correlation between ploidy level and complexity as well as the geographical distribution of organisms with different ploidy levels.

## 4.0 The advantage of being haploid

The theory developed in the previous sections attempts to explain why diploidy is advantageous. However, haploidy must also have advantages otherwise no organisms would be haploid. Even prokaryotes would be diploid (or higher) if there were no trade-offs that decreased the fitness of higher ploid organisms. Two such potential trade-offs are genetic load and growth rate.

Haldane (1937) argued that the genetic load in a population is directly proportional to the mutation rate. Thus, a species with twice the mutation rate of another species will have twice the genetic load within its population. If mutation rates are relatively constant at all ploidy levels, then diploid organisms should have twice the genetic load of haploid organisms (Otto and Whitten, 2000, Gerstein and Otto, 2009). As a consequence, populations of haploid organisms should have higher mean fitnesses than populations of otherwise identical diploid organisms (Orr and Otto, 1994).

Higher ploidy levels are also disadvantageous because they lead to slower growth rates. Polyploid organisms typically grow and mature more slowly than their diploid relatives (Otto and Whitten, 2000; Hessen *et al.*, 2009). In one species of bryophyte, diploid gametophyte lines grow ~70% as fast as haploid lines on full medium (Schween *et al.*, 2005). Fern sporophytes produced by apogamy (growth of unfertilized eggs) are haploid and mature faster than diploid sporophytes produced via fertilization (Sharpe and Mehltreter, 2010). Perhaps lower ploidy levels are favored when the advantages of rapid growth outweigh the advantages of lower protein self-binding. Thus, haploidy would be particularly beneficial to single-celled organisms, such as bacteria, which proliferate rapidly via cell division.

The growth rate hypothesis is particularly useful when trying to rationalize the life cycles of plants. All plants must compete for limited space, so a faster growth rate would give haploid plants an advantage over diploid plants when attempting to claim an area of land. For instance, bryophytes reproduce asexually via fragmentation and sexually via spore production. In both cases, the gametophyte plants must quickly grow from only a few cells in order to establish themselves in a piece of land. Given that bryophytes are the plants that benefit the least from heterozygosity, they would benefit more from the faster growth rates associated with being haploid. Ferns also disperse themselves via spores made up of only a few cells. Thus, their simple haploid gametophyte generations may be beneficial because they can quickly establish themselves in a piece of land. Then, the ferns reproduce sexually and produce their complex diploid sporophyte generations, which don't have to compete for access to land because they grow out of their gametophyte parents. Spermatophytes disperse themselves via seeds that carry entire diploid plant embryos. The plant embryos can quickly establish themselves in a piece of land, despite a slower growth rate, because they are already partially developed prior to germination. This might allow spermatophytes to be diploid complex organisms for the bulk of their life-cycle, which in turn might allow them to utilize their complexity for all of their life processes, such as sexual reproduction.

**5.0 Problem Organisms**

Some organisms do not fit neatly within the theoretical framework presented here. This section contains a brief description of these problem organisms and highlights the difficulties they pose for the theory presented in this paper.

*5.1 Isomorphic algae*

Some species of algae, such as those in the genus *Ulva*, alternate between isomorphic gametophyte and sporophyte generations. Such species are a problem for the theory because neither generation is more complex than the other, and therefore, both generations should express proteins of similar length. Every generation should be diploid if the proteins are sufficiently long, and every generation should be haploid if the proteins are sufficiently short. However, alternating between haploid and diploid generations may allow the algae to deal with stress. Mohsen *et al.* (1973) showed that temperatures below 20°C favored growth of gametophytes while temperatures between 20-30°C favored sporophytes. Perhaps sporophytes are favored at warm temperatures because their proteins are less prone to aggregation. Experimental evidence has also shown that *Ulva* are more sensitive to desiccation stress at warmer temperatures than at cooler temperatures (Zou *et al.*, 2007). *Ulva* increase their expression of *Hsp70* as a response to desiccation stress, confirming that desiccation does promote protein aggregation in these species (Fu *et al.*, 2010). Lee (2008) argues that *Ulva* grow in mid to upper inter-tidal zones during cooler months and are restricted to lower inter-tidal zones during summer months because they are more sensitive to desiccation at warm temperatures, though no data is provided. Members of another genus of inter-tidal macroalgae, *Porphyra*, are also sensitive to temperature. *Porphyra* grow as haploid macroalgae in the winter months and survive the summer months as diploid microscopic filaments that burrow inside mollusk shells (Hoek *et al.*, 1995). Thus, *Ulva* and *Porphyra* may grow as haploid organisms in the less stressful months of the year because they can grow and mature faster as haploids, and then they may switch to a diploid stage of their life-cycle during the stressful months.

However, observations such as these are likely to depend on the species. For instance, Alström-Rapaport *et al.* (2010) found that the population of *Ulva* in the Baltic Sea, which is ice-covered in the winter, is dominated by sporophytes. The proportion of gametophytes in the population increases from 10% to 35% over the course of the summer and then drops off as the weather cools. The imbalance is maintained because many of the sporophytes in the Baltic Sea reproduce asexually by producing diploid spores that over-winter. Also, Hiraoka and Yoshida (2010) found that *Ulva* in Hiroshima Bay, Japan, exhibited no seasonal fluctuations between sporophytes and gametophytes. They observed that the generations alternated yearly. Thus, while there is some evidence that *Ulva* sporophytes are hardier than gametophytes, there is no clear evidence that sporophyte hardiness influences the life-cycle of *Ulva*.

The relative metabolic efficiencies of the sporophyte and gametophyte generations may also influence *Ulva* and *Porphyra* life-cycles. Sporophytes should be more metabolically efficient than gametophytes because their heterozygosity would result in lower protein turnover rates (see Appendix 1). This may be beneficial in low-nitrogen environments. Mohsen *et al.* (1974), for example, showed that low nitrogen concentrations induced gametophytes to release their gametes, which led to the growth of sporophytes in their culture medium.

*5.2 Diploid single-celled organisms*

Several species of diploid single-celled organisms exist. Among them are ciliates, plasmodial slime molds, oomycetes, diatoms, and a few algal species. Such organisms are compatible with the theory presented in this paper because diploidy should be more strongly correlated with protein length than with organism complexity. These species may represent examples of morphologically simple organisms that synthesize long, aggregation-prone proteins. Interestingly, many of the diploid single-celled organisms (such as diatoms, oomycetes, opalinids, and raphidophytes) are stramenopiles. Since these organisms are genetically related, they may synthesize similar proteins. Perhaps many stramenopile species synthesize long, multi-domain proteins for reasons unrelated to morphological complexity. The diploid-dominant kelps, which are the largest and most complex of the seaweeds, are also stramenopiles, so morphological complexity does occur within this taxonomic group. These species provide another test for the theory presented in this paper. The correlation between diploidy and long proteins should be stronger than the correlation between diploidy and morphological complexity. Therefore, a systematic study comparing the lengths of proteins among various species should include both morphologically complex and simple diploid species to verify that diploidy is indeed correlated with protein length and not morphological complexity.

**6.0 Conclusion**

This paper has proposed that the ploidy level of a species is controlled by the lengths of proteins synthesized by its members. The argument is:

1. Long proteins are more susceptible to forming soluble oligomers than short proteins because their folding rate constants are lower and their self-binding rate constants are higher.
2. The rate of a protein's self-binding reaction can be slowed by diluting the concentration of the protein.
3. Two different allozymes will be synthesized by an organism that is heterozygous at the allozymes' gene locus. Each allozyme will have a lower concentration within the heterozygous organism than it would if the organism were homozygous at that gene locus.
4. From (1)-(3), heterozygosity is beneficial to organisms that synthesize many long, multi-domain proteins, and such organisms will tend to have higher ploidy levels.
5. Organisms that synthesize shorter proteins have higher fitnesses when their ploidy levels are lower because lower ploidy levels lead to faster growth rates, shorter maturation times, and lower genetic loads.
6. Complex organisms tend to synthesize longer proteins than simple organisms.
7. From (1)-(6), complex organisms tend to be diploid because they tend to synthesize long proteins, and simple organisms tend to be haploid because they tend to synthesize short proteins.
8. Organisms that synthesize long proteins suffer from high levels of protein self-binding in harsh environments that promote protein aggregation, and are less likely to survive and adapt in such environments.
9. Polyploids are pre-adapted to harsh environments because they have high heterozygosities and because they have more than two different alleles at many of their gene loci.
10. Organisms that synthesize short proteins are also pre-adapted to harsh environments.
11. From (6)-(10), harsh environments that promote protein aggregation tend to be inhabited by simple organisms and polyploid complex organisms.

The general conclusion of this paper is that many of the benefits of heterozygosity are statistical mechanical rather than genetic. The main value of this theory is its ability to explain the link between complexity and ploidy level as well as the geographical distributions associated with organisms of each ploidy level.

**Acknowledgments:** I would like to thank Jeffry Mitton for his recommendations to improve this paper.

**Appendix A: Selection for heterozygosity**

*Epistasis and truncation selection*

Lewontin and Hubby (1966) provided an influential argument against the hypothesis that natural selection favors heterozygous genotypes. They used allozyme data to estimate the genetic diversity within different *Drosophila Pseudoobscura* populations and found that the average individual heterozygosity ranged from 8-15% of gene loci, depending on the sampled population. Furthermore, they found that protein polymorphisms were segregating at approximately one third of all gene loci within each population, which would mean that approximately 2,000 gene loci are polymorphic within *D. pseudoobscura* populations. If it is assumed that homozygosity at a single gene locus reduces the reproductive potential of an individual by 10%, and that only two polymorphisms are segregating at the gene locus at a frequency of 50% each, then half of the individuals within the population will be homozygous at the gene locus, and the reproductive potential of the whole population will be reduced by 5%. This would mean that that the reproductive fitness of the population would be $0.95^{2000}$, or $10^{-46}$, its maximum value. This is an unrealistically low number, and they concluded that natural selection could not favor heterozygotes. Other studies confirmed that numerous polymorphisms are segregating in wild populations, and a debate ensued between biologists who thought the high levels of genetic diversity were favored by natural selection and biologists who thought the diversity was neutral (Kimura, 1968; Crow, 1992).

King (1967), Milkman (1967), and Sved *et al.* (1967) proposed that high levels of genetic diversity could be maintained by truncation selection without an unreasonable cost to reproductive potential. In a truncation selection model, all individuals whose heterozygosities are below a critical value have a fitness of zero, and individuals whose heterozygosities are above the critical value have maximal fitness. Figure 3 shows a truncation model in which 2,000 gene loci are polymorphic, and exactly two polymorphisms exist at each gene locus with a frequency of 50% each. In such a population, the average heterozygosity of an individual will be 1,000, the variance will be 500, and 5% of the population will be heterozygous at less than 956 gene loci. Wills (1978) showed that truncation of the bottom 5% of a population can maintain polymorphisms at 66,000 gene loci in a population of 100,000 individuals. The effectiveness of truncation selection comes from its severity and its ability to operate on many gene loci simultaneously. Several studies have found evidence for truncation selection acting on wild populations, but the results have been mixed (see Mitton, 1997 for review and Kaeuffer *et al.*, 2007 as a recent example).

Truncation selection can be considered an extreme form of positive epistasis in which homozygosity confers increasingly severe fitness costs with each additional homozygous gene locus. Several authors have shown that molecular chaperones can facilitate epistasis by inhibiting protein aggregation (Fares *et al.*, 2002; Sollars *et al.*, 2003; Maisnier-Patin *et al.*, 2005; de Visser and Elena, 2007; de Visser *et al.*, 2011; Lehner, 2011). These papers assume that deleterious mutations code for unstable, aggregation prone proteins, and that organisms carrying many deleterious mutations must increase their expression of molecular chaperones in order to inhibit protein aggregation. However, if an organism carries too many deleterious mutations, then its molecular chaperones may be overwhelmed and the organism's fitness may be substantially lowered.

This section develops two simple models that link an organism's heterozygosity to its fitness. Like the main text, heterozygosity at a gene locus is assumed to decrease the concentrations of the polymorphisms that each allele encodes. It is also assumed that deleterious mutations are not responsible for the fitness costs of homozygosity. Another assumption is that the rate of soluble oligomer disassociation is trivial compared to the rate of soluble oligomer formation. This latter assumption is justified because soluble oligomers accumulate within organisms if they do not use molecular chaperones to remove them (McClellen *et al.*, 2005). Finally, it is assumed that a single molecular chaperone is responsible for removing soluble oligomers.

First, every soluble oligomer that forms within an organism must be removed. A steady-state concentration of soluble oligomer can be maintained if the rate of soluble oligomer formation, *dA/dt*, equals the rate of removal, *dR/dt*:

$$\frac{dA}{dt} = \frac{dR}{dt} \tag{A1}$$

The rate of removal can be expressed in terms of the Michaelis-Menton Equation for the molecular chaperone (Kondepudi, 2008):

$$\frac{dR}{dt} = \frac{R_{max}[O]_{steady}}{K_m + [O]_{steady}} \tag{A2}$$

where $R_{max}$ is the chaperone's maximum rate of removal, $[O]_{steady}$ is the steady-state concentration of soluble oligomer, and $K_m$ is the Michaelis-Menton constant for the molecular chaperone. The rate of addition is (from Equation 8):

$$\frac{dA}{dt} = \sum_j \sum_i k_{bji} r_{ji}^2 [U_{ji}]^2 \tag{A3}$$

where $[U_{ji}]$ is the concentration of unfolded polypeptide chain expressed by each allele *i* at each gene locus *j*, $r_{ji}$ is the concentration of unfolded polypeptide chain *ji* in an organism that is heterozygous for *ji* divided by its concentration in an organism that is homozygous for *ji*, and $k_{bji}$ is the rate law constant for the self-binding reaction of each unfolded polypeptide chain encoded by each allele at each gene locus. In order to keep the model simple, the rest of this appendix will assume that every allele at every gene locus encodes polypeptide chains with the same values of $k_b$ and $[U]_{steady}$ under any particular circumstances. The rest of the appendix will also assume that $r_i$ is equal to one for homozygous gene loci and 0.5 for heterozygous loci. These simplifying assumptions give:

$$\frac{dA}{dt} = k_1(N_{Hom} + 0.5 N_{Het}) = k_1(T - 0.5 N_{Het}) \tag{A4}$$

where $N_{Hom}$ is the number of homozygous gene loci, $N_{het}$ is the number of heterozygous gene loci, *T* is the total number of polymorphic gene loci, and $k_1 = k_b[U]^2_{steady}$. The parameter $k_1$ quantifies the propensity of an organism's proteins to form soluble oligomers. It increases with the stressfulness of the environment and the length of the organism's proteins.

Combining Equations A1-A4 and solving for $[O]_{steady}$ gives:

$$[O]_{steady} = \frac{K_m k_1 (T - 0.5 N_{Het})}{R_{max} - k_1 (T - 0.5 N_{Het})} \tag{A5}$$

The steady-state concentration of soluble oligomer increases as d$A$/d$t$ approaches the value of $R_{max}$. The molecular chaperone is overwhelmed when d$A$/d$t$ is greater than $R_{max}$, and soluble oligomer accumulates faster than the molecular chaperone can remove it. Thus, there can be no steady-state concentration when $dA/dt$ is greater than $R_{max}$ and Equation A5 is meaningless.

Soluble oligomers are toxic and associated with many diseases, so an organism's fitness should decline with higher concentrations of soluble oligomer (Kayed *et al.*, 2003; Haas and Selkoe, 2007; Viera *et al.*, 2007). The relative fitness, *w*, of the organisms can be represented by:

$$w = 1 - \alpha [O]_{steady} \qquad (A6)$$

where the value of *w* is relative to the fitness of an organism that produces no soluble oligomers and $\alpha$ is the reciprocal of the lethal concentration of soluble oligomer. Combining Equations A5 and A6 gives:

$$w = 1 - \frac{\alpha K_m k_1 (T - 0.5 N_{Het})}{R_{max} - k_1 (T - 0.5 N_{Het})} \qquad (A7)$$

In this model, the relative fitness of the organism drops to zero as $dA/dt$ approaches $R_{max}$ because the molecular chaperone is increasingly taxed by higher rates of soluble oligomer formation, which leads to higher steady-state concentrations of soluble oligomer.

Figure 4 shows the results of Equation A7 for different values of $\alpha K_m$. High values of $\alpha K_m$ generate curves with gradual declines in relative fitness while low values of $\alpha K_m$ generate curves with steep declines similar to truncation selection. The value of $K_m$ affects the steepness of the decline because it quantifies the efficacy of the molecular chaperone. A low $K_m$ value means that the molecular chaperone can maintain high rates of soluble oligomer removal until $dA/dt$ approximately equals $R_{max}$, then the molecular chaperone is overwhelmed and relative fitness collapses. The value of $\alpha$ affects the steepness of the decline because it quantifies how much relative fitness declines with each small increase in $[O]_{steady}$.

Figure 4 also shows the results of Equation A7 for different values of $k_1$. Increasing the value of $k_1$ increases the heterozygosity at which truncation occurs. This is interesting because $k_1$ quantifies the propensity of proteins to form soluble oligomers, which should increase with the stressfulness of the environment. Thus, Equation A7 predicts that only highly heterozygous individuals will be found in stressful environments because less heterozygous individuals will be to the left of the truncation line. The main text cites several studies showing that polyploids tend to be associated with stressful environments that promote protein aggregation. Polyploids can possess more than two alleles at each gene locus, which would reduce their total rate of soluble oligomer formation, as described in Equation A3. Furthermore, they have high heterozygosities because most are hybrid species (Otto and Whitton, 2000). Thus, a higher proportion of organisms may be polyploid in stressful environments because truncation selection removes organisms with lower ploidy levels. Additionally, polyploidy and hybridization may have fitness costs (such as high genetic loads, slower growth rates, and outbreeding depression), so polyploid organisms may be restricted to stressful environments where they do not have to compete with diploid organisms (Otto and Whitton, 2000).

Figure 4 can also illustrate the relationship between protein length and ploidy level described in the main text. The value of $k_1$ should increase with the lengths of the proteins synthesized by an organism. As a consequence, organisms that synthesize short enough proteins can be haploid because they should not be to the left of a truncation line even though they are homozygous at all gene loci. In contrast, organisms that synthesize long enough proteins will have to be heterozygous at some gene loci in order to stay to the right of a truncation line, so they will have to be (at least) diploid.

The value of $k_1$ increases with both the length of an organism's proteins and the stressfulness of the environment. Hence, organisms that synthesize long proteins, such as animals, should have extremely high $k_1$ values in harsh environments that promote protein aggregation. This will result in truncated fitness curves that prevent such organisms from migrating into and adapting to harsh environments. In contrast, organisms that synthesize very short proteins, such as bacteria and archaea, should still have small $k_1$ values in harsh environments. Such organisms can migrate into harsh environments because their fitness does not plummet in harsh environments. As a result, harsh environments should mostly be inhabited by simple organisms and a few complex organisms (Longton, 1988; Alpert, 2006). These patterns will be less clear cut for organisms that synthesize intermediate length proteins. For example, bryophytes are haploid-dominant and grow in harsh environments that cannot support more complex plants (Longton, 1988). However, there are diploid and triploid varieties of bryophytes, which occur with greater frequency at higher latitudes (Wyett *et al.*, 1988; Ricci *et al.*, 2008). Thus, bryophytes probably synthesize short enough proteins that they can survive as haploid organisms in benign environments, but their proteins are probably long enough that they require heterozygosity to survive in harsh environments.

Equation A7 assumes that heterozygosity comes with no cost, but this may not be true. One of the polymorphisms encoded at a gene locus may be inferior to the other and confer a fitness cost to its possessor. The issue may be addressed by modifying Equation A6:

$$w = \left(1 - \frac{\alpha K_m k_1 (T - 0.5 N_{Het})}{R_{max} - k_1 (T - 0.5 N_{Het})}\right)(1 - hs)^{N_{Het}} (1 - s)^{q(T - N_{Het})} \tag{A8}$$

where the value of $w$ is relative to the fitness of an organism that doesn't possess any suboptimal polymorphisms and doesn't produce any soluble oligomers, $q$ is the percentage of homozygous gene loci carrying inferior alleles, $s$ is the fitness cost of the inferior alleles, and $h$ is the dominance of the inferior alleles (h=1 is completely dominant and h=0 is completely recessive). The equation obviously assumes each inferior allele has the same fitness cost and dominance. Figure 5 shows the results of Equation A8 for different values of *hs*. The value of *hs* must be low or else the relative fitness of the organism will be low. For instance, the relative fitness of the organism never exceeds 0.35 when *s* equals 0.001, *h* equals 0.5, and *q* equals 0.5, but the relative fitness of the organism has a maximum value of 0.82 when the value of *s* equals 0.0002.

The model developed in this appendix is meant to provide a simplified framework for understanding how protein self-binding can lead to truncation selection that maintains a large number of molecular polymorphisms in natural populations. The model attempts to show that an organism's defenses against protein aggregation can be overwhelmed if the rates of protein self-binding are sufficiently high, and this is more likely to occur in organisms with low heterozygosities. The model is testable because it clearly predicts that genetic diversity should increase with the stressfulness of the environment. The increasing abundance of hybrid polyploid species in harsh environments seems to support the prediction (see main text). In contrast, the neutral theory of molecular evolution predicts that genetic diversity should increase with effective population size (Tajima, 1983). The currently available evidence doesn't seem to support this prediction (Bazin *et al.*, 2006; Hahn, 2008; Leffler *et al.*, 2012). Therefore, the model presented in this appendix should at least be considered when attempting to explain genetic variation in natural populations.

*Inbreeding depression and heterosis*

Equations A5-A8 hold the concentration of molecular chaperone constant and allow the steady-state concentration of soluble oligomers to change. However, the reverse approach can also be taken to develop an alternative relationship between relative fitness and heterozygosity. In this model, it will be assumed that the organism maintains soluble oligomers at a critical steady-state concentration. If the steady-state concentration of the soluble oligomers rises, then the concentration of the molecular chaperone will increase to lower the soluble oligomers' steady-state concentration back to their critical level. This may be accomplished through a feedback mechanism, such as the unfolded protein response (UPR) that occurs inside the endoplasmic reticulum (Shroeder et al., 2005; Bernales et al., 2006). The model will also assume that the organism maintains steady-state concentrations of functional proteins that are continuously unfolding, forming soluble oligomers, and being replaced by new proteins. The resulting model will link the protein turnover in an organism to its heterozygosity.

First, the value of $R_{max}$ is directly proportional to the concentration of molecular chaperone:

$$R_{max} = k_2 M_t \tag{A9}$$

where $M_t$ is the total concentration of molecular chaperone and $k_2$ is the rate law constant for the molecular chaperone's catalyzing reaction. Combining Equations A5 and A9 and solving for $M_t$ gives:

$$M_t = k_1 [T - 0.5 N_{Het}] \left[ \frac{K_m + [O]_{steady}}{k_2 [O]_{steady}} \right] \tag{A10}$$

Equation A10 gives the concentration of molecular chaperone necessary to maintain a particular steady-state concentration of soluble oligomers at a given heterozygosity. It predicts that inbred organisms will have higher concentrations of molecular chaperones than outbred organisms, which has been confirmed by Kristensen *et al.* (2002). If a molecular chaperone marks an unfolded polypeptide chain for degradation by tagging it with ubiquitin, then it will be destroyed by either the proteosome or the lysozome (Hayes and Dice, 1996, Kopito, 2000; McClellen *et al.*, 2005, Rubinsteinz, 2006). The rate of soluble oligomer removal should be proportional to the number of molecular chaperones:

$$\frac{dR}{dt} = \frac{k_2 M_t [O]_{steady}}{K_m + [O]_{steady}} = k_1 [T - 0.5 N_{Het}] \tag{A11}$$

The rate of soluble oligomer removal is equal to the rate of soluble oligomer addition under steady-state conditions. The degradation of soluble oligomers is an ATP consuming process, and the rate of calories consumed by their degradation should be in proportion to the rate that they are degraded:

$$\frac{dC_{Degrad}}{dt} = \beta \frac{dR}{dt} \tag{A12}$$

where $dC_{Degrad}/dt$ is the rate of calories consumed due to soluble oligomer degradation and $\beta$ is the number of calories consumed during the degradation of each marked soluble oligomer. Combining Equations A10-A12 gives:

$$\frac{dC_{Degrad}}{dt} = \beta k_1 [T - 0.5N_{Het}] \tag{A13}$$

In addition, the degradation of soluble oligomers must be balanced by the synthesis of new proteins in order to maintain a steady-state concentration of properly functioning protein within the cytosol:

$$\frac{dP}{dt} = \eta \frac{dA}{dt} = \eta k_1 [T - 0.5N_{Het}] \tag{A14}$$

where $dP/dt$ is the rate that new proteins are synthesized to replace proteins that have formed soluble oligomers and $\eta$ is the number of proteins making up the soluble oligomers. Equations A11 and A14 predict that inbred organisms should have higher rates of protein turnover than outbred organisms, which has been confirmed by Hawkins *et al.* (1986). The synthesis of proteins is an ATP consuming process, and the rate of calories consumed by their synthesis should be in proportion to the rate that they are synthesized:

$$\frac{dC_{Synth}}{dt} = \gamma \frac{dP}{dt} = \gamma \eta k_1 [T - 0.5N_{Het}] \tag{A15}$$

where $dC_{Synth}/dt$ is the rate of calories consumed due to protein replacement and $\gamma$ is the number of calories consumed during the synthesis of each replacement protein. The overall calorie consumption rate due to protein maintenance is:

$$\frac{dC_{Maint}}{dt} = \frac{dC_{Degrad}}{dt} + \frac{dC_{Synth}}{dt} \tag{A16}$$

where $dC_{Maint}/dt$ is the calorie consumption rate due to protein maintenance. Combining Equations A11-A16 gives:

$$\frac{dC_{Maint}}{dt} = k_1 [\beta + \gamma \eta][T - 0.5N_{Het}] \tag{A17}$$

Equation A17 predicts that inbred organisms should be less metabolically efficient than outbred organisms, which has been confirmed by many studies (see Mitton, 1997 for review). The relative fitness of the organism should decline as its caloric maintenance costs increase. Hence:

$$w = 1 - \delta \frac{dC_{Maint}}{dt} \tag{A18}$$

where the value of $w$ is relative to the fitness of an organism that has no maintenance costs and $\delta$ is the reciprocal of the lethal maintenance costs. Combining Equations A17 and A18 gives:

$$w = 1 - k_1 \delta [\beta + \gamma \eta][T - 0.5N_{Het}] \tag{A19}$$

In Equation A19, the relative fitness of an organism increases linearly with heterozygosity because the organism must expend a smaller fraction of its available calories to maintain a specific steady-state concentration of soluble oligomer and native protein.

Studies of inbreeding depression and heterosis usually compare the performance of inbred organisms relative to outbreds. Therefore, Equation A19 should be expressed in relative terms:

$$w_{relative} = \frac{w_{inbred}}{w_{outbred}} = \frac{1-\kappa k_1[T-0.5 N_{Het}^{inbred}]}{1-\kappa k_1[T-0.5 N_{Het}^{outbred}]} \tag{A20}$$

where $w_{inbred}$ and $w_{outbred}$ are the relative fitnesses of the inbred and outbred organisms, $N_{Het}^{inbred}$ and $N_{Het}^{outbred}$ are the number of heterozygous gene loci in the inbred and outbred organisms, and $\kappa$ aggregates the Greek constants in Equation A19. Equations A9-A20 provide a theoretical model for the relationship between protein aggregation, inbreeding depression, and heterosis. The model is in agreement with previous research that found a correlation between heterozygosity and: (1) metabolically efficiency (Mitton, 1997), (2) protein turnover (Hawkins *et al.*, 1986), and (3) expression of molecular chaperones (Kristensen *et al.*, 2002). Figure 6 shows the results of Equation A20 for different values of $k_1$, which increases with the stressfulness of the environment. As can be seen, Equation A20 predicts that inbreeding depression should increase in stressful environments, which is also supported by previous research (Armbruster and Reed, 2005).

Equations A7 and A8 differ substantially from Equations A19 and A20. They represent two different extreme scenarios in which an organism either cannot increase its concentration of molecular chaperone or can increase its concentration of molecular chaperone without limit. The true physiological behavior of organisms is probably somewhere in between these extremes. Organisms can probably increase their concentrations of molecular chaperones up to certain maximum amounts. In all likelihood, their relative fitness would decline linearly with homozygosity until the environment reaches a certain stressful limit. Then, the relative fitness will decline severely with homozygosity, resulting in truncation selection. The rate of stress increase will also have an impact. Organisms may not be able to increase their concentrations of molecular chaperones rapidly enough to respond to a sudden stressful event, such as a hard freeze. In fact, several studies have shown that organisms respond to severe stresses better after a period of mild stress exposure that allows them to increase their expression of molecular chaperones (Clark and Worland, 2008; Janská *et al.*, 2010). Thus, Equations A19 and A20 are probably most applicable to organisms exposed to slowly increasing and mild stresses while Equations A7 and A8 are probably most applicable to organisms exposed to rapidly increasing and severe stresses.

*Alternation of generations*

The main text argued that ferns alternate between a simple, haploid gametophyte generation that grows quickly and a complex, diploid sporophyte generation that probably produces longer, more aggregation-prone proteins. This hypothesis is supported by the ability of fern gametophytes to tolerate stresses that fern sporophytes cannot. Similarly, bryophyte sporophytes are less tolerant to stresses than bryophyte gametophytes (Stark *et al.*, 2007), so the sporophytes may also produce longer proteins than the gametophytes. However, this hypothesis cannot work for all organisms that alternate between haploid and diploid life-cycle stages. For example, *Ulva* alternates between isomorphic haploid and diploid generations that probably produce proteins of similar lengths. In addition, many other species of algae and fungi enter into a temporary diploid (or dikaryotic) stage prior to the formation of resting spores. The proteins expressed during these organisms' sporulation may be longer than the proteins expressed during the rest of the organisms' life-cycles, but this is unlikely.

Instead, diploidy may allow some organisms to produce spores that are more tolerant of environmental stresses. For example, ascomycetes commonly produce two types of spores: conidia, which are produced by haploid parents, and ascospores, which are produced by dikaryotic parents. Ascospores are more resistant to stresses that promote protein aggregation than conidia (Dijksterhuis, 2007). In fact, ascospores make up a larger percentage of all spores produced by ascomycetes as the harshness of the environment increases (Grishkan *et al.*, 2003). Therefore, ascospores may be able to survive in environments that conidia cannot because of truncation selection. A truncation selection model could also explain why some organisms do not produce resting spores unless they enter into a diploid state. For example, diploid strains of the yeast *Saccharomyces cerevisiae* produce spores, but unmated haploid strains do not. Perhaps *S. cerevisiae* never evolved the ability to produce spores from haploid parents because such spores would be to the left of a truncation line, such as those shown in Figure 4, and would have zero fitness. Therefore, truncation selection may explain why many eukaryotes, not just fungi, enter into a diploid or dikaryotic state prior to forming resting spores.

This hypothesis has two complications. First, many resting spores, including ascospores, are haploid. However, they are formed after the meiotic division of diploid cells, so they may inherit the mixed-up cytoplasm of their parents. Second, the spores would have to be metabolically active in order for the truncation selection model to work because the model assumes that molecular chaperones remove soluble oligomers when they form (Equation A1). Spores are generally considered dormant, but some researchers have reported evidence that yeast spores are metabolically active, at least under some conditions (Barton *et al.*, 1982; Brengues *et al.*, 2002). These metabolically active spores can lose their viability over a period of time, presumably because they run out of energy (Brengues *et al.*, 2002).

Combining the ideas presented in this section with those of the main text can potentially explain the evolution of different life-cycles. They are (in order of their probable appearance on Earth): 1) Prokaryotes produce short proteins and their life-cycles do not contain any diploid stages. 2) Some eukaryotes, such as some algae and fungi, may produce proteins that are sufficiently short for their haploid stages to not be to the left of a truncation line in benign environments, but their proteins may be sufficiently long that diploidy is necessary to survive in harsh environments. Such organisms should alternate between haploid and diploid life-cycle stages as the harshness of the environment fluctuates. Their haploid stages are often vegetative and their diploid stages are often either resting spores (or cysts) or the parents of resting spores. 3) Other eukaryotes, such as bryophytes and ferns, may alternate between haploid and diploid generations that produce proteins of different lengths. 4) Finally, some eukaryotes, such as animals and spermatophytes, may produce sufficiently long proteins that they are primarily diploid, even in benign environments.

Two additional observations concerning the evolution of life-cycles can be made. First, if the truncation model presented above is true, then haploid dominant life-cycles with temporary diploid stages (2 above) may be necessary precursors to the evolution diploid dominant life-cycles (4 above). This is because truncation selection would inhibit the evolution of long, aggregation-prone proteins in strictly haploid organisms. Second, truncation selection wouldn't favor a temporary diploid stage unless it is already heterozygous when it appears. Thus, temporary diploid stages may have evolved in single-celled organisms when cells from closely related species fused together to provide the heterozygosity necessary to survive in harsh environments. This would be analogous to the formation of highly heterozygous polyploid species via hybridization in more complex organisms.

**Appendix B: Heterozygosity, chemical potential, and chemical affinity**

Equations A7, A8, A19 and A20 assume that organisms remove soluble oligomers from their cytosol as quickly as they form, which leads to steady-state concentrations of the soluble oligomers. Equations A19 and A20 also assume organisms maintain steady state concentrations of native proteins because the proteins are continuously unfolding, forming soluble oligomers, and are being replaced by new functional proteins. The chemical potentials of a properly folded (native) protein polymorphism, $P$, in organisms that are homozygous and heterozygous for $P$ are:

$$\mu_{native} = \mu°_{native} + kTln[N_P]_{steady} \quad \text{and} \quad \mu_{native} = \mu°_{native} + kTln[N_P]_{steady} + kTln(r_{native}) \quad (B1)$$

where $\mu_{native}$ is the chemical potential of native $P$, $\mu°_{native}$ is the standard state chemical potential of native $P$, $k$ is Boltzmann's constant, $T$ is temperature, $[N_P]_{steady}$ is the steady state concentration of native $P$ in an organism that is homozygous for $P$, and $r_{native}$ is the steady-state concentration of native $P$ in an organism that is heterozygous for $P$ divided by $[N_P]_{steady}$. The equations show that the chemical potential of a native polymorphism is $kTln(r_{native})$ lower in an organism that is heterozygous for the polymorphism than in an organism that is homozygous for the polymorphism. The chemical potentials of a soluble oligomer of $P$ will be the same as in Equation B1 for organisms that are homozygous and heterozygous for $P$.

The thermodynamic driving force of a chemical reaction is its chemical affinity divided by temperature. Chemical affinity, $A$, is the negative of the derivative of Gibbs free energy, $G$, with respect to a reaction progress variable, $\xi$ (Kondepudi, 2008):

$$A \equiv -\left(\frac{dG}{d\xi}\right) \quad (B2)$$

It follows from this definition that chemical affinity is also equal to the difference between the chemical potentials of the reactants and the chemical potentials of the products. The chemical affinity for the formation of a soluble oligomer of $P$ in an organism that is homozyous for $P$ is:

$$A_{Hom} = 2\mu_{native} - \mu_{oligomer} = 2\mu°_{native} - \mu°_{oligomer} + 2kTln[N_P]_{steady} - kTln[O]_{steady} \quad (B3)$$

where $A_{Hom}$ is the chemical affinity of $P$'s self-binding reaction, $\mu_{oligomer}$ is the chemical potential of the soluble oligomer, and $[O]_{steady}$ is the steady-state concentration of the soluble oligomer. The chemical affinity for the formation of a soluble oligomer of $P$ in an organism that is heterozygous for $P$ is:

$$A_{Het} = 2\mu°_{native} - \mu°_{oligomer} + 2kTln[N_P]_{steady} - kTln[O]_{steady} + 2kTln(r_{native}) - kTln(r_{oligomer}) \quad (B4)$$

where $A_{Het}$ is the chemical affinity of $P$'s self-binding reaction, $[O]_{steady}$ is the steady state concentration of the soluble oligomer in an organism that is homozygous for $P$, and $r_{oligomer}$ is the steady-state concentration of the soluble oligomer in an organism that is heterozygous for $P$ divided by $[O]_{steady}$. The difference in chemical affinity between organisms that are homozygous and heterozygous for $P$ is:

$$\frac{A_{Hom} - A_{Het}}{T} = kln(r_{oligomer}) - 2kln(r_{native}) \quad (B5)$$

Equation B5 can be interpreted in two equivalent ways. First, for a given gradient of steady-state concentrations between native *P* and a soluble oligomer of *P*, an organism that is homozygous for *P* will be further away from chemical equilibrium than an organisms that is heterozygous for *P* by the amount given in Equation B5. Second, the thermodynamic push of the self-binding reaction is weaker in an organism that is heterozygous for *P* than in an organism that is homozygous for *P* by the amount given in Equation B5.

Unfolded polypeptide chains of *P* are continuously formed by the unfolding of native *P*, and they are continuously removed by the refolding of *P* and by the formation of soluble oligomers. The assumed steady-state concentrations of native *P* and soluble oligomers of *P* imply that the concentration of unfolded *P* also has a steady-state concentration. The relative chemical potentials of unfolded *P* will be the same as in Equation B1 for organisms that are homozygous and heterozygous for *P*. This will lead to a modified version of Equation B5 for the formation of soluble oligomers of *P* from unfolded *P*:

$$\frac{A_{Hom} - A_{Het}}{T} = k\ln(r_{oligomer}) - 2k\ln(r_{unfolded}) \tag{B6}$$

where $r_{unfolded}$ is the steady-state concentration of unfolded *P* in an organism that is heterozygous for *P* divided by $[O]_{steady}$. Chemical affinity can be related to the rates of elementary step reactions by (Kondepudi, 2008):

$$\frac{A}{T} = k\ln\left(\frac{R_f}{R_r}\right) \tag{B7}$$

where $R_f$ is the rate of the forward reaction (the rate of protein self-binding) and $R_r$ is the rate of the reverse reaction (the rate of soluble oligomer disassociation). Combining Equations B6 and B7 gives:

$$\left(\frac{R_f}{R_r}\right)_{Hom} = \left(\frac{r_{oligomer}}{r_{unfolded}^2}\right)\left(\frac{R_f}{R_r}\right)_{Het} \tag{B8}$$

Equation B8 implies that the forward and reverse reactions for the formation of a polymorphism's soluble oligomer will be more biased in favor of the forward reaction in an organism that is homozygous for the polymorphism than in an organism that is heterozygous for the polymorphism.

A chemical reaction's rate of progression is given by:

$$\frac{d\xi}{dt} = R_f - R_r \tag{B9}$$

where ξ is the reaction progress variable. Combining Equations B7 and B9 gives:

$$\frac{d\xi}{dt} = R_f\left(1 - e^{\frac{-A}{kT}}\right) \tag{B10}$$

Equation B10 shows that chemical reactions proceed towards equilibrium when their chemical affinity does not equal zero. Thus, organisms must expend energy to remove soluble oligomers because they maintain steady-state concentrations of native protein, unfolded polypeptide chain, and soluble oligomer that are different from their equilibrium values.

Equation B10 also shows that the rate of chemical progression increases with a chemical system's distance from chemical equilibrium. Hence, inbred organisms have higher rates of calorie expenditure than outbred organisms because they maintain steady-state concentrations of native protein, unfolded polypeptide chain, and soluble oligomer that are farther away from chemical equilibrium. As such, living organisms are behaving like other physical systems that maintain non-equilibrium gradients. For example, a refrigerator consumes more power when its internal temperature is farther away from the the temperature of its surroundings because heat flow within a system increases with the system's temperature gradient. Equation B10 shows why the theory of heterosis presented in this paper is statistical mechanical. The protein turnovers and metabolic efficiencies of organisms wouldn't be affected by the organisms' heterozygosities if the organisms weren't maintaining non-equilibrium biochemistries. Likewise, severe environments kill organisms because they cause organisms' biochemistries to move toward equilibrium faster than the organisms can restore their internal steady-states. The constant movements toward and away from chemical equilibrium are responsible for organisms' maintenance costs, and give rise to the differences in performance between inbred and outbred organisms.

748.

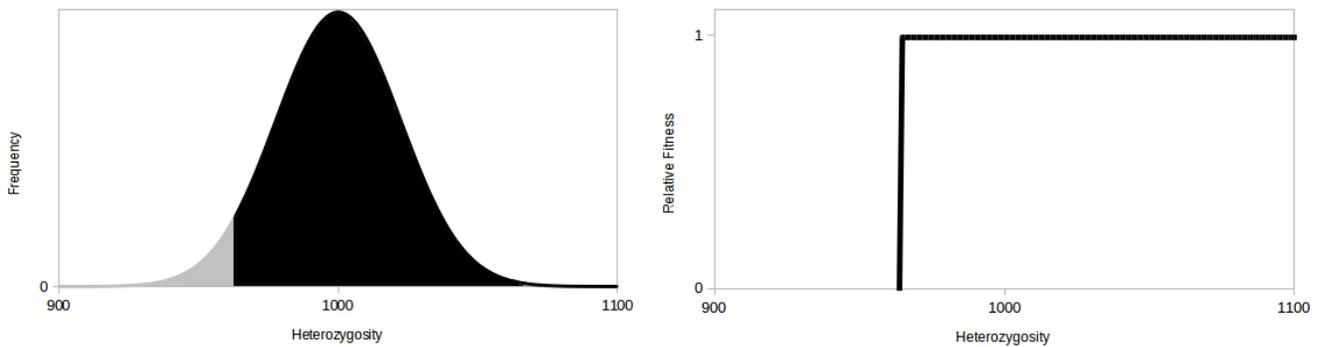

Figure 3. Conceptual model for truncation selection. The left figure shows the distribution of heterozygosities in a population. The gray area represents the bottom 5% of the population. The right figure shows the relative fitness of an organism as a function of heterozygosity. All organisms with heterozygosities in the bottom 5% have zero relative fitness.

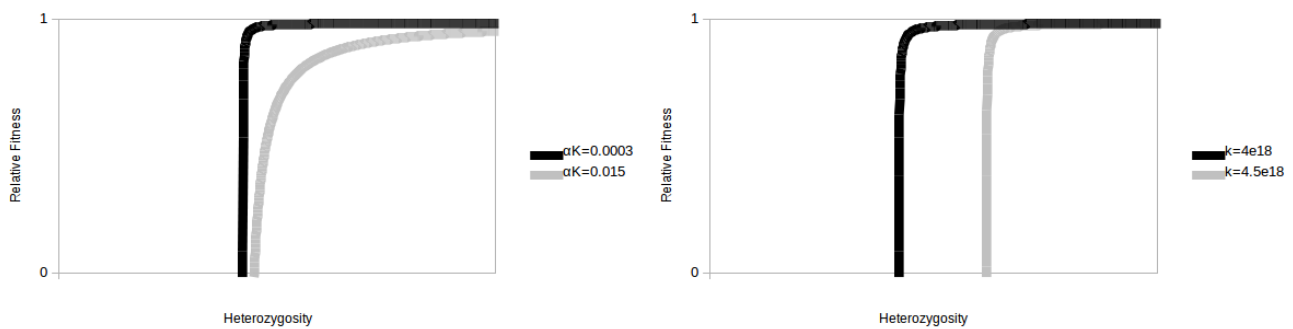

Figure 4. Results from Equation A7. The left diagram compares results for two different values of $\alpha K_m$. Higher values of $\alpha K_m$ (gray curve) result in more gradual declines in fitness. The right diagram compares results for two different values of $k_1$. Higher values of $k_1$ (gray curve) result in truncation at a higher heterozygosity.

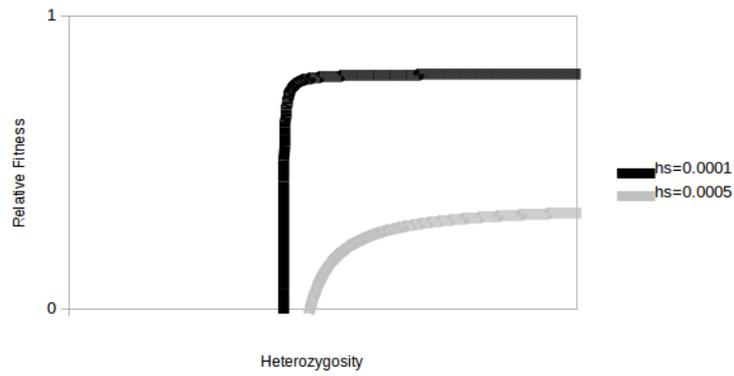

Figure 5. Results from Equation A8. Fitness declines with increasing heterozygosity because one of the polymorphisms at each gene locus is slightly deleterious. The gray curve shows the results for a higher *hs* value.

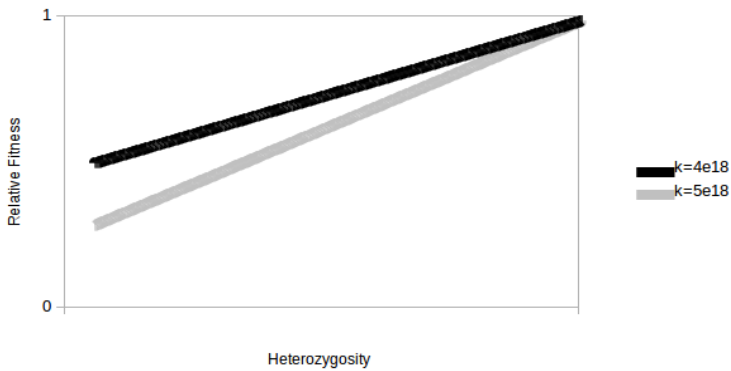

Figure 6. Results from Equation A20. Relative fitness increases linearly with heterozygosity. The gray curve shows the results for a higher $k_1$ value, which increases with both the stressfulness of the environment and the length of the proteins synthesized by the organism.